# Performing Homomorphic Encryption Experiments on IBM's Cloud Quantum Computing Platform

He-Liang Huang,[1,2] You-Wei Zhao,[2,3] Tan Li,[1,2] Feng-Guang Li,[1,2] Yu-Tao Du,[1,2] Xiang-Qun Fu,[1,2] Shuo Zhang,[1,2] Xiang Wang,[1,2] and Wan-Su Bao[1,2,*]

[1]*Zhengzhou Information Science and Technology Institute, Henan, Zhengzhou 450000, China*
[2]*CAS Centre for Excellence and Synergetic Innovation Centre in Quantum Information and Quantum Physics, University of Science and Technology of China, Hefei, Anhui 230026, China*
[3]*Hefei National Laboratory for Physical Sciences at Microscale and Department of Modern Physics, University of Science and Technology of China, Hefei, Anhui 230026, China*
(Dated: November 8, 2016)

Quantum computing has undergone rapid development in recent years. Due to the limitations on scalability, personal quantum computers still seem a little unrealistic in near future. The first practical quantum computer for ordinary users is likely to be on the cloud. However, the adoption of the cloud computing is applicable only if security is ensured. Holomorphic encryption is cryptographic protocol that allows computation to be performed on encrypted data without decrypting them, which is therefore well suited for cloud computing. Here we first applied the holomorphic encryption on IBM's cloud quantum computer platform. Our experiments successfully implemented the quantum algorithm for linear equations with our privacy being protected. This demonstration opens up a feasible way to the next stage of the development in cloud quantum information technology.

## 1. Inroduction

In recent years, much progress has been made in developing quantum computing technologies [1, 2]. Because of the quantum superposition principle, quantum computers could outperform their classical counterparts when solving certain problems. For example, Shor's algorithm [3–7], quantum simulation [8–11], solving linear systems of equations [12–14], quantum machine learning [15, 16], and so on. Therefore, the emergence of quantum computers will change the world once again. Due to its high construction costs and maintenance costs, the first quantum computer is likely to be only owned by a small number of organizations. Fortunately, however, with the cloud service, ordinary users can also be able to apply to use these quantum computers.

As expected, recently IBM has made a 5-qubit quantum computer publicly available over the cloud [17]. Based on a 5-qubit superconducting chip in a star geometry, and a full Clifford algebra, the system can be reprogrammable and allowed for circuit design, simulation. Through the classical internet, users can easily test and execute the quantum algorithms on interactive platform called *QuantumExperience* (QE). So far, several experiments have already been reported [18–20].

The future cloud quantum computing is likely to be available to users as the way similar to the IBM's cloud quantum computing platform. That is, the interaction between users and IBM's cloud computing platform is achieved through the website. In this case, the quantum circuit, input data and output data of users are completely public to the server. While sharing the computational resources of quantum computing on the cloud, we also need to consider the privacy. Although there are a number of protocols and experiments have been proposed to develop the secure cloud quantum computing [21–24], but these encryption method do not suit for the current level of technology, since the input data or output data cannot be public to the servers on the website in the previous protocol.

In the world of classical cryptography, homomorphic encryption [25–27] is scheme which allows certain operations to be performed on the encrypted data without decryption. Thus, users can provide encrypted data to a remote server for processing without having to reveal the plaintext. Although the data is open to the server, but the server cannot reveal the real data, since the data is encrypted. After the server output the results to users, users can recover actual output data through his privacy key. Therefore, homomorphic encryption has become a practical encryption technique for cloud computing.

In this paper, we designed a homomorphic encryption protocol for the cloud quantum computing, which is suitable for IBM's cloud server. Based on the basic quantum gates provided by the server, we developed a series construction methods for various operations. Finally, we successfully implemented the quantum algorithm for linear equations on the IBM's cloud server with our privacy being protected. This work will hopefully motivate more people to get involved in this field, since this is the first time to consider the security of users data on the IBM's cloud server, and can provide a guidance for future large-scale cloud quantum computing.

## 2. Methods

To solve linear equations on quantum computer, we employ the quantum algorithm proposed by Harrow *et al.* [12], which can provide an exponential speedup over existing classical algorithms. Given a matrix $A$ and a vector $\vec{b}$, we aim to solve the equations $A\vec{x} = \vec{b}$. To adapt the problem to quantum version, we rescale $\vec{x}$ and $\vec{b}$ to $||\vec{x}|| = ||\vec{b}|| = 1$. Thus we can encode the problem as

$$A|x\rangle = |b\rangle \tag{1}$$

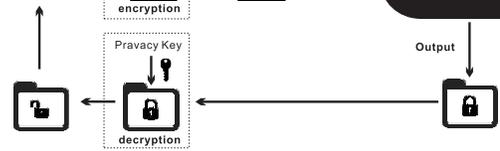

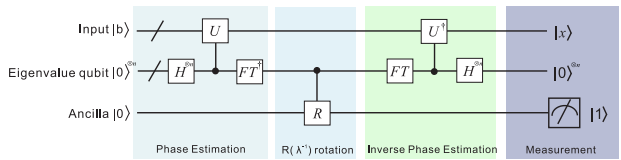

FIG. 1. (color online). Quantum circuit for solving systems of linear equations. The circuit of the original quantum algorithm. In the circuit, $U = \sum_{k=0}^{T-1} |k\rangle\langle k| \otimes e^{iAkt_0/T}$, $H$ is the Hadamard gate. $FT$ is the Fourier transformation and $FT^\dagger$ is the inverse Fourier transformation. The output $|x\rangle$ is obtained when the ancilla is detected as $|1\rangle$.

Denote $\{|u_i\rangle\}$ and $\{\lambda_i\}$ as the eigenbasis and eigenvalues of matrix $A$. The input state $|b\rangle$ can be expanded in the eigenbasis of $A$ as $|b\rangle = \sum_{i=1}^{N} \beta_i |u_i\rangle$, where $\beta_i = \langle u_i|b\rangle$. To seek the solution $|x\rangle = A^{-1}|b\rangle/||A^{-1}|b\rangle||$, the algorithm can be decomposed into three main steps (see Fig. 1).

In the first step, phase estimation is applied to the transform $|b\rangle|0\rangle^{\otimes n}$ into $\sum_{i=1}^{N} \beta_i |u_i\rangle|\lambda_i\rangle$, where $|0\rangle^{\otimes n}$ is the eigenvalue register of $n$ qubits, and eigenvalues $|\lambda_i\rangle$ are stored in eigenvalue register after phase estimation.

In the second step, one needs to implement the map $|\lambda_i\rangle \to \lambda_i^{-1}|\lambda_i\rangle$ to extract the eigenvalues of $A^{-1}$. By implementing a controlled $R(\lambda^{-1})$ rotation on an additional ancilla qubit intitially in the state $|0\rangle$, we can transforms the system to

$$\sum_{i=1}^{N} \beta_i |u_i\rangle|\lambda_i\rangle \left( \sqrt{1 - \frac{C^2}{\lambda_i^2}}|0\rangle + \frac{C}{\lambda_i}|1\rangle \right) \tag{2}$$

The final step is to implement the inverse phase estimation to disentangle the eigenvalue register to $|0\rangle^{\otimes n}$. Then, we end up with

$$\sum_{i=1}^{N} \beta_i |u_i\rangle \left( \sqrt{1 - \frac{C^2}{\lambda_i^2}}|0\rangle + \frac{C}{\lambda_i}|1\rangle \right) \tag{3}$$

After measuring and postselecting the ancillary qubit of $|1\rangle$, we will obtain an output state $\sum_{i=1}^{N} C(\beta_i/\lambda_i)|u_i\rangle$, which is proportional to our expected results state $|x\rangle$.

To delegate the task of solving linear equations to IBM's cloud quantum computing platform, one can directly encode the quantum circuit on the servers without considering security. This is equivalent to send the cloud server the matrix $A$ and vector $\vec{b}$ directly. However, when facing the issue of security, this approach becomes no longer feasible. Inspired by the homomorphic encryption (see Fig. 2), we can compile a homomorphic encryption version of the algorithm.

FIG. 2. (color online). The homomorphic encryption scheme. User encrypt the data using privacy key before sending it to the cloud quantum server. Server do not own the privacy key so that it cannot learn anything about the encrypted data. If the operations of server are homomorphic operations, then server can utilize arbitrary computations on the encrypted data without decrypting it. The output of the server remains in encrypted form, and can only be recovered by the user who have the privacy key.

To implement the homomorphic encryption, we must ensure that the operations of the server and the encryption scheme of user meet the following conditions,

$$f(E(x)) = E(f(x)) \tag{4}$$

Where $f$ denotes the operations of the server, and $E(x)$ is used to denote the encryption of the message $x$. According to the conditions of homomorphic encryption, we can design the protocol as follows:

Step 1. For a linear equation $A\vec{x} = \vec{b}$ with $n$ variants, user ranodomly chooses $a_i \in \{0,1\}$ as private keys, and let $x_i = y_i + a_i$, then substitutes it into the linear equations

$$A(\vec{y}_i + \vec{a}_i) = \vec{b} \tag{5}$$

User rewrite the equations as $A\vec{y}_i = \vec{b}'$, where $\vec{b}' = \vec{b} - A\vec{a}_i$. Then send the $A$ and $\vec{b}'$ to cloud server.

Step 2. Server implement the quantum algorithm to solving the equations received from user, and feedback the results $\vec{r}$ to user.

Step3. User decrypts the results as

$$x_i = r_i + a_i \tag{6}$$

Here, we analyze the whole process of the protocol. Step 1 can be regard as the encryption process. That is, for $\forall \vec{x}$, $E(\vec{x}) = \vec{x}'$, where $\vec{x}'$ is the encryption message of $\vec{x}$. Let $g(\vec{x}')$ and $F$ be the linear equations and the algorithm for solving linear equations, respectively. Then step 2 can be represented as $F(g(\vec{x}')) \to \vec{x}'$. Finally, step 3 is used to decrypt the results as $D(\vec{x}') \to \vec{x}$, where $D$ denotes the operations of decryption, then the user can obtained actual results. Note that with the processing of homomorphic encryption, the input and output of user are perfectly hidden and the server cannot get any of the private data, since all the server does is to deal with encrypted data.

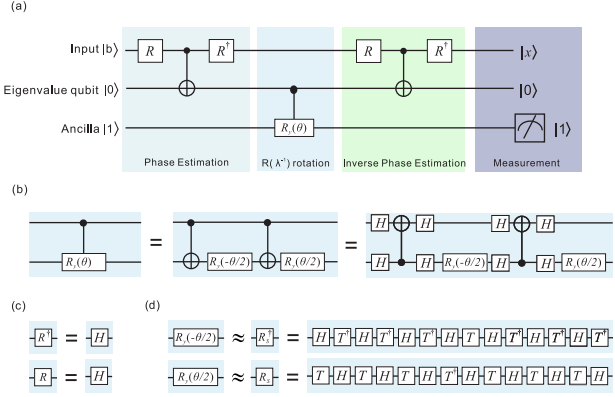

FIG. 3. (color online). The construction of the gates for solving the equations (9) and (10). (a) The optimised circuit for $2 \times 2$ system of linear equations. $R$ is a unitary operation that can diagonalizes the matrix $A$ as $A = R^\dagger \begin{pmatrix} \lambda_1 & 0 \\ 0 & \lambda_2 \end{pmatrix} R$, where $\lambda_i$ is the eigenvalue of $A$. (b) The construction of the controlled $R_y(\theta)$. (c) The R gate in (a) can be compiled to Hadamard gate for solving the equations (9) and (10). (d) The construction of the $R_y(\theta/2)$ and $R_y(-\theta/2)$ in the (b).

### 3. Experimental Realization

Here we demonstrate a proof-of-principle experiment of this protocol on IBM cloud quantum computing platform. Using one state qubit as the two-vector $|b\rangle$, one eigenvalue qubit and ancilla qubit, we can utilize the protocol to solving systems of $2 \times 2$ linear equations. The quantum circuit of the algorithm for $2 \times 2$ linear equations can be compiled into the circuit shown as Fig. 3(a) [13]. A unitary $R$ is introduced to diagonalizes the matrix $A$ as $A = R^\dagger \begin{pmatrix} \lambda_1 & 0 \\ 0 & \lambda_2 \end{pmatrix} R$, where $\lambda_i$ is the eigenvalue of $A$. $R(\lambda^{-1})$ rotation can be realized by a controlled $R_y(\theta)$, where $R_y(\theta) = \exp(-i\theta\sigma_y/2)$, $\sigma_y$ is the usual Pauli matrix, and $\theta$ is controlled by the eigenvalue qubit with the function $\theta_i = -2\arccos(\lambda_1/\lambda_2)$. The algorithm succeeds with probability when ancilla qubit is measured in the state $|1\rangle$. In our implementation, we choose two systems of linear equations to be

$$\begin{pmatrix} 0.7 & 0.3 \\ 0.3 & 0.7 \end{pmatrix} \cdot \vec{x} = \begin{pmatrix} 1/\sqrt{2} + 0.7 \\ 1/\sqrt{2} + 0.3 \end{pmatrix} \quad (7)$$

$$\begin{pmatrix} 1.75 & 0.75 \\ 0.75 & 1.75 \end{pmatrix} \cdot \vec{x} = \begin{pmatrix} 1/\sqrt{2} + 1.75 \\ -1/\sqrt{2} + 0.75 \end{pmatrix} \quad (8)$$

Without loss of generality, we set the private key of user are $a_1 = 1, a_2 = 0$. By substituting $x_i = y_i + a_i$ into the linear equations, user can perfectly hide his input data, and the equations can be rewritten as

$$\begin{pmatrix} 0.7 & 0.3 \\ 0.3 & 0.7 \end{pmatrix} \cdot \vec{y} = \begin{pmatrix} 1 \\ 1 \end{pmatrix} / \sqrt{2} \quad (9)$$

$$\begin{pmatrix} 1.75 & 0.75 \\ 0.75 & 1.75 \end{pmatrix} \cdot \vec{y} = \begin{pmatrix} 1 \\ -1 \end{pmatrix} / \sqrt{2} \quad (10)$$

Then user can encode the circuit on IBM's cloud quantum computing platform. For both of these two linear equations, the $R$ gate in the Fig. 3(a) can be compiled into Hadamard gate (see Fig. 3(c)). Note that IBM only provide the CNOT gate as two-qubit gate, to realize the controlled $R_y(\theta)$ operation ($\theta$ is equal to $-57.34°$ for both of the two linear equations), we decomposed the controlled $R_y(\theta)$ gate into two CNOT gates, one $R_y(\theta/2)$ and one $R_y(-\theta/2)$ gate. In our implementation, we set the eigenvalue register as the central qubit of the IBM's superconductor quantum chip. As the chip only allows to operate CNOT gates with the central qubit as target qubit in their star geometry. If we want to operate CNOT gates with the central qubit as control qubit, we need to combine a CNOT and four Hadamard gates to achieve this task. Then the controlled $R_y(\theta)$ gate can be compiled to the combination of several Hadamard gates, CNOT gates, $R_y(\theta/2)$ and $R_y(-\theta/2)$ gate (see Fig. 3(b)). Now the question shifts to how to construct an R gate, since that only Clifford gates ($X, Y, Z, H, S, S^\dagger$ and CNOT) and two non-Clifford gates ($T$ and $T^\dagger$) are available on the platform. It turns out that adding almost any non-Clifford gate to the Clifford gates is universal [28]. Therefore, by adding the $T$ gate in Clifford gates, it is possible to reach all different points of the Bloch sphere. Monte Carlo simulation indicates that the more $T$ gates in our circuit, we can cover the Bloch sphere more densely with states we can reach. Figure 4 depicts the attainable states by adding at most 1, 3, 5, 7 $T$ gates to the Clifford gates respectively.

In the Fig. 4, the red dot is the $R_y(\theta/2)$ operation we desired. Based on the results of numerical simulation, we can approximate the $R_y(\theta/2)$ gate with the gate $R_S$, which is the combination of seven $T$ gates and seven Hadamard gates (see Fig. 3(d)). To characterize its accuracy, we compute the similarity $F = 1/2 \cdot Tr(U_{ideal} U_{simu})$ as 0.998, where $U_{ideal}$ is the ideal unitary operation $R_y(\theta/2)$ and $U_{simu}$ is the simulated unitary operation $R_S$, indicating that our simulated unitary operation is very similar to the ideal operation $R_y(\theta/2)$. Through the red dot ($R_y(\theta/2)$ operation) and purple dot (simulated operation $R_S$) in Fig. 4(d), it is more intuitive that these two dots are very close. At this point, we can compile the full circuit for solving the two equations on the IBM servers.

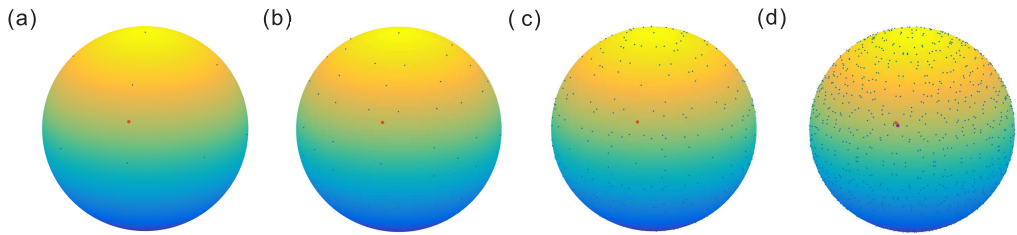

FIG. 4. (color online). (a), (b), (c) and (d) are the Bloch sphere with the dots are the attainable states of $U|0\rangle$, where $U$ is the operation by adding at most 1, 3, 5 and 7 $T$ gates to the Clifford gates respectively. The red dot in (a), (b), (c) and (d) is the $R_y(\theta/2)$ operation we desired for solving the equations (9) and (10). The purple dot in (d) is the simulated operation of $R_y(\theta/2)$.

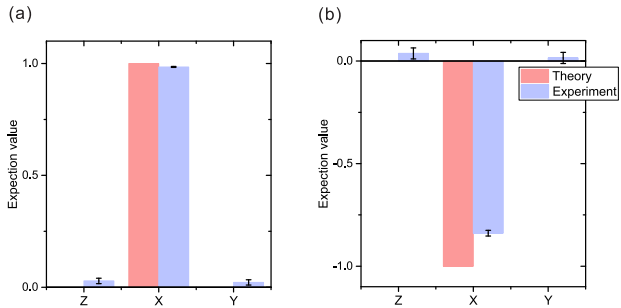

FIG. 5. (color online). Experimental results. (a) and (b) are the measurement results of the output state of equations (9) and (10). For each equations, the ideal (red bar) and experimentally obtained (blue bar) expectation values of the Pauli Z, X, and Y are presented. The error bars denote one standard deviation, deduced from propagated Poissonian counting statistics of the raw detection events.

### 4. Results

Measuring the first qubit of the circuit in Fig. 3(a) in Pauli $Z, X, Y$ basis, we can obtain the solutions of equations. Figure 5(a) and Figure 5(b) shows that both the ideal (red bar) and experimentally obtained (blue bar) expectation values for each Pauli operator when implementing the algorithm to solve equations (9) and (10). We compute the fidelity of output state as $F = \langle x|\rho_{\exp}|x\rangle$, where $|x\rangle$ is the ideal output state and $\rho_{\exp}$ is the experimentally output state from the measurement results of Pauli $Z, X, Y$. The output states have fidelities of 0.992(1) and 0.920(7) for equations (9) and (10) respectively, indicating a high reliable results in our experiments.

By post-processing the results using classical computer, user can easily decrypt the secret results to obtain actual results as $\{x_1 = 1.7173, x_2 = 0.6967\}$ and $\{x_1 = 1.7227, x_2 = 0.6911\}$ equations (7) and (8), respectively. Theoretical analysis shows that the error is within 2% compared with the actual solution. Then, the protocol of homomorphic encryption is announced successful.

### 5. Conclusion

In summary, we have presented the first experimental demonstration of the homomorphic encryption protocol for solving linear equations on IBM's cloud quantum computer platform. The protocol is very suitable for the current technology, which enables the users delegate the task of computation by encoding the circuit on website of the quantum servers with their data being protected. Even though current quantum computations in IBM's server are proof-of-principle demonstrations, it can be scalable for larger systems in the future. Ideally, this work will provide a workable solution to the future cloud quantum computation.

### Acknowledge

The author acknowledges the use of IBM's Quantum Experience for this work. The views expressed are those of the author and do not reflect the official policy or position of IBM or the IBM Quantum Experience team. This project is supported by the National Basic Research Program of China (Grant No.2013CB338002), National Natural Science Foundation of China(Grants No.11504430 and No.61502526).